\documentclass[conference]{IEEEtran}
\IEEEoverridecommandlockouts
\usepackage{dblfloatfix}
\usepackage{tabularx}
\usepackage{wrapfig}
\usepackage{amsmath}
\usepackage{color}
\usepackage{amsmath,amsfonts,amssymb,amscd}
\usepackage{booktabs}
\usepackage{paralist}
\usepackage{enumitem}
\usepackage{multicol}
\usepackage{gensymb}
\usepackage{steinmetz}
\usepackage{mathtools}
\usepackage{siunitx}
\usepackage{hyperref}
\usepackage{cuted}

\usepackage{mathrsfs}
\usepackage{multirow}
\usepackage{multirow}
\usepackage{flushend}
\usepackage{cleveref}
\usepackage{graphicx}
\usepackage{subcaption}
\usepackage{xurl}

\begin{document}

\title{Impact Assessment of Data Integrity Attacks in MVDC Shipboard Power Systems}

\author{
\IEEEauthorblockN{\textbf{Kirti~Gupta}\IEEEauthorrefmark{1}, \textbf{Subham Sahoo}\IEEEauthorrefmark{2},  \textbf{Bijaya Ketan Panigrahi}\IEEEauthorrefmark{1},  \textbf{and Charalambos Konstantinou}\IEEEauthorrefmark{3}}

\IEEEauthorblockA{
\IEEEauthorrefmark{1}Department of Electrical Engineering, Indian Institute of Technology -- Delhi, Delhi, 110016, India\\
\IEEEauthorrefmark{2}Department of Energy, Aalborg University, Aalborg, 9220, Denmark\\
\IEEEauthorrefmark{3}CEMSE Division, King Abdullah University of Science and Technology (KAUST), Thuwal 23955-6900, Saudi Arabia
}

\IEEEauthorblockA{
E-mail: \{Kirti.Gupta, Bijaya.Ketan.Panigrahi\}@ee.iitd.ac.in,} sssa@energy.aau.dk, charalambos.konstantinou@kaust.edu.sa
}

\IEEEaftertitletext{\vspace{0\baselineskip}}

\maketitle
\begin{abstract} 
The development of power electronics-based medium voltage direct current (MVDC) networks has revolutionized the marine industry by enabling all-electric ships (AES). This technology facilitates the integration of heterogeneous resources and improves efficiency. The independent shipboard power system (SPS) is controlled by exchanging measurements and control signals over a communication network. However, the reliance on communication channels raises concerns about the potential exploitation of vulnerabilities leading to cyber-attacks that could disrupt the system. In this paper, a notional 12 kV MVDC SPS model with zonal electrical distribution system (ZEDS) architecture is considered as an exemplary model. As the system stability is closely linked to the transient performance, we investigate how to determine the operational status of the system under potential data integrity attacks on the governor and exciter of the power generation modules (PGMs). Further, the impact of these attacks on the stability of rotor speed and the DC link voltage is derived and discussed. The simulation of the system is carried out in MATLAB/Simulink environment. 
\end{abstract}


\IEEEpeerreviewmaketitle

\section{Introduction}
The advancements in medium voltage switch components is a primary contributor to the electrification of ships to become all electric ship (AES). This transformation improved power density, enhanced reliability, aided reduction in the weight of onboard electrical components, and hence, boosted flexibility \cite{1}. On contrary to the terrestrial distribution systems, the marine distribution system faces several challenges like close electrical and mechanical proximity, limited generation units, stringent operating conditions etc. Consequently, shipboard power systems (SPS) are typically more prone to faults and failures, and thus, there is a requirement of effective reconfiguration strategies for enhanced efficiency and management of electrical power during adverse situations \cite{2}. In this work, the zonal electrical distribution system (ZEDS) is considered, whose functional and control aspects are presented in IEEE Std 1826$^{\mathrm{TM}}$-2012 \cite{3} and IEEE Std 1676$^{\mathrm{TM}}$-2010 \cite{4}. The whole system is partitioned in zones to facilitate controlled power exchanges. It also has the capability to reconfigure itself to supply the critical loads continuously either from port or starboard buses in the adverse conditions. 

In order to facilitate controlled generation and management of power, the whole SPS is interconnected through communication channels, supported by GPS, {electronic chart display information systems (ECDIS)}, {automatic identification systems (AISs)}, etc., which play a vital role in the proper operation of SPS. Thus, the resulting cyber-physical SPS becomes potentially vulnerable to cyber-attacks. The cyber-threat analysis within maritime sector is of paramount importance. It became the only industry whose four major companies suffered from cyber-intrusions in recent years \cite{5}. These include the NotPetya ransomware attack on \textit{APM-Maersk} in 2017, which caused service interruption for weeks; the ransomware attack on \textit{COSCO} in 2018; the malware attack in the \textit{Mediterranean Shipping Company} in 2020, which brought down the data center for days; and the Rangar Locker ransomware attack on \textit{CMA CGM}, which took down booking system in 2020. In addition, other incidents include a {GPS jamming attack} in South Korea affected 280 vessels in 2016 \cite{6}; a {ransomware attack} in 2021 on shipping companies \cite{7}; and a {malicious code installation} in 2022 providing unauthorized access \cite{8}. Among several types of cyber-attacks \cite{9}, this work focuses on data integrity attacks which aim to modify measurement and control information transmitted over communication media. In particular, our work analyzes the impact of false data injection attacks (FDIAs) which falsify data via injecting information to cause misleading of operational system algorithms \cite{27}.

The recent literature contributes on the control of SPS to improve its operational efficiency \cite{10}. In \cite{11}, the authors presented the modeling and control of the hybrid DC SPS. Further, the multiagent-based management of power to enhance efficiency is discussed in \cite{12}. The effect of communication networks as well as their modeling in a real-time simulation environment is elaborated in \cite{ogilvie2020modeling, 9512343} with the purpose of evaluating controls of SPS. In the field of cybersecurity for SPS, review is conducted in \cite{13}, which only presents a brief overview of a list of vulnerabilities present in SPS. In this paper, we focus on the impact of data integrity attack on SPS. Data integrity attacks can also impact the stability of the system \cite{14}. The SPS stability illustrates the ability of the SPS to re-attain a stable operating point after being subjected to large disturbance (e.g., faults, FDIAs, etc.). Such large FDIAs can cause disconnection of generator/loads leading to local/full shutdown \cite{9512317}, hence affecting the reliability of electrical supply. To the best of the authors' knowledge, this paper is first of its kind to discuss the impact of data integrity attacks on the stability of MVDC SPS. To inspect the stability of the SPS, we investigate the transient response of the rotor speed and DC link voltage of the exemplary system in this paper.

This paper considers a 12 kV nominal two-zone SPS with ZEDS. The impact of data integrity attacks on the power generation modules (PGMs) causing instability in the system is studied. A general FDIA model is presented including an amplification term, a constant term, and a time-varying term. Further, the closed-form expressions of the rotor angle and speed as well as the DC link voltage dynamics are discussed for the considered FDIAs model. The key contributions of this work can be summarized as follows:
\begin{itemize}
    \item The data integrity attacks are demonstrated on the governor and excitation system of the synchronous generators (part of PGMs) of MVDC SPS. The attack model encapsulates the fabrication of rotor speed and DC link voltage measurements to cause tripping of the installed relays.
    \item The transient analysis on rotor speed and DC link voltage is derived and discussed for such attacks.
\end{itemize}

The remainder of the paper is organized as follows. A brief description of the vulnerabilities in the MVDC SPS is presented in the Section II. The transient analysis of the rotor speed and DC link voltage dynamics is  derived and discussed for data integrity attacks on governor and exciter in Section III. The results and discussions are presented in Section IV. Finally, the concluding remarks are illustrated in Section V.

\section{Vulnerability Assessment of MVDC SPS}

\begin{figure}[t]
    \centering
    \includegraphics[width=0.5\textwidth]{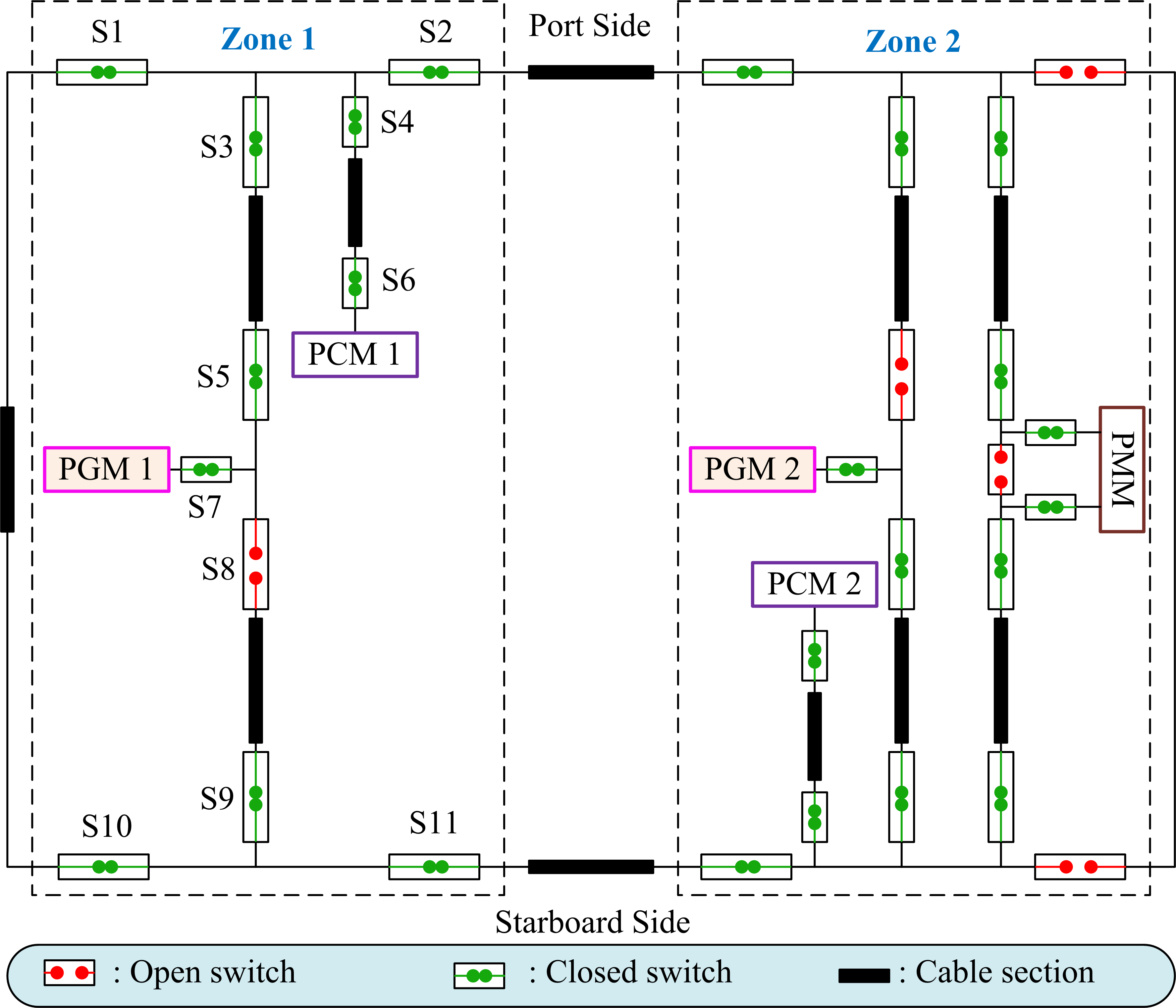}
    \caption{Notional two-zone MVDC SPS.}
    \label{fig:Two}
\end{figure}

The SPS model considered in this work is the `{Notional Two- Zone MVDC Power System}', developed under the Electric Ship Research and Development Consortium (ESRDC) and shown in Fig. \ref{fig:Two}. The parameters of the various elements in this model can be referred from \cite{15}. Here, each zone comprises of PGMs, power conversion modules (PCMs), propulsion motor modules (PMMs) connected by cable sections through switches (S). In zone 1, the switches are mentioned from S1 to S11. Similarly, it can be defined in zone 2. Since the control of such a complex system is highly dependent on the communication network, the system reliability and stability may be threatened from cyber-attacks. 

Some of the vulnerabilities of modern SPS include \cite{16}: \textit{(i)} falsified sharing of information, malware injections, spoofing, jamming attacks on {AIS} causing hijacking of ships, data theft etc.; \textit{(ii)} fabricated data injections in {ECDIS} due to presence of obsolete operating systems resulting in hijacking, data theft, compromised computers, etc.; \textit{(iii)} jamming attacks, spoofing, (distributed) denial of service (DoS)/(DDoS), packet modification in {GPS} causing hijacking of ships, false information, disruption and delay of services. Although several entry points of attacks could be potentially utilized in MVDC SPS, this work mainly focuses on the data integrity attacks on the PGMs. A PGM comprises of synchronous generator and the converter along with their control modules. In particular, we present the effect of FDIAs on the governor and exciter of the synchronous generators. The attack model for each case is described below.

\subsection{Governor Attack Model}
\begin{figure}[t]
    \centering
    \includegraphics[width=0.85\columnwidth]{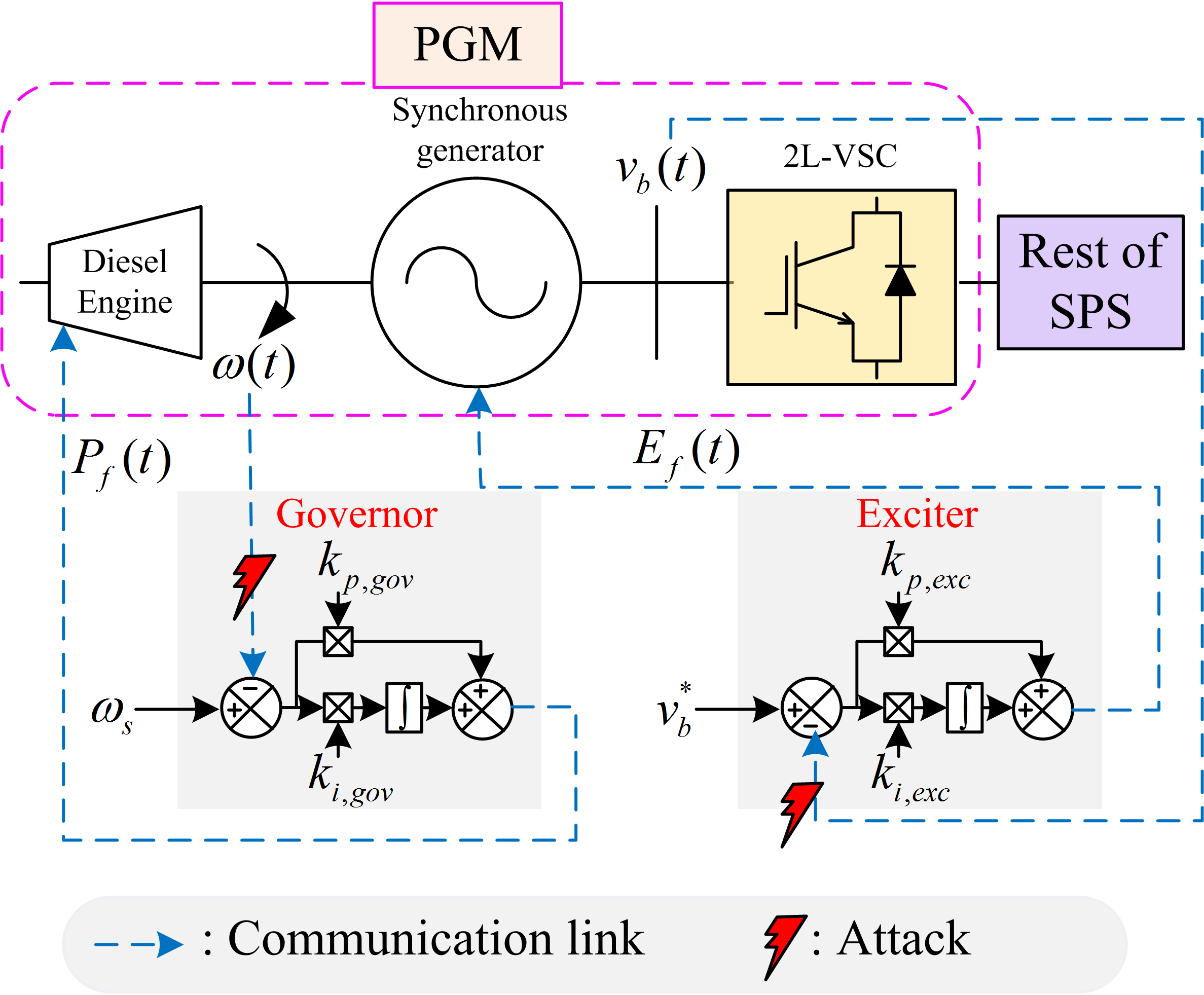}
    \caption{Synchronous generator control architecture.}
    \label{fig:Sync}
\end{figure}

The governor is responsible to regulate the generator speed according to the variation in loads, consequently maintaining the system frequency to the synchronous speed ($\omega_s$), as shown in Fig. \ref{fig:Sync}. The dynamics of the governor ($x_{gov}$) can be represented as \cite{17}:
\begin{equation}
\label{eq1}
\left\{ \begin{aligned}
  & \dot{x}_{gov}(t)=\omega_s-\omega_i(t) \\ 
 & P_f(t)=k_{p,gov_i}\left ( \omega_s-\omega_i(t) \right )+k_{i,gov_i}x_{gov,i}(t) \\ 
\end{aligned} \right.
\end{equation}

\noindent where, the subscript $i$ denotes the parameters corresponding to the $i^{th}$ synchronous generator, $\omega$ is the measured rotor speed, $P_f$ is the fuel index, $k_{p,gov}$ is the proportional gain and $k_{i,gov}$ is the integral gain. The control, as in any other cyber-physical system, can be realized through digital control schemes linked over communication channels as presented in Fig. \ref{fig:Sync}, via various protocols like IEC 61850 over Ethernet interface \cite{18}, CAN \cite{19}, TCP/IP \cite{20}, etc. The vulnerabilities of these protocols are described in \cite{21,22}, elaborating the ways to interrupt the confidentiality, integrity and availability of the messages over these protocols. In our work, it is assumed that the attacker can obtain the information of rotor speed measurements and later fabricate these measurements to cause anomalous situations.

\subsection{Exciter Attack Model}
The excitation system controls the generator field voltage, responsible to control the source voltage generated from the synchronous generator (Fig. \ref{fig:Sync}). The dynamics of the exciter ($x_{exc}$) can be expressed as \cite{17}:
\begin{equation}
\label{eq2}
    \dot{x}_{exc,i}(t)=v_{b,i}^*-v_{b,i}(t)
\end{equation}
where, $v_{b,i}(t)=\sqrt{v_{bd,i}^2(t)+v_{bq,i}^2(t)}$ represents the measured bus voltage. The reference bus voltage is represented by $v_b^*$. The terms `$d$' and `$q$' represent the $d-$axis and $q-$axis quantities, respectively. The field voltage ($E_f$) is expressed as follows:
\begin{equation}
\label{eq3}
    E_{f,i}(t)=k_{p,exc_i}\left ( v_{b,i}^*-v_{b,i}(t) \right )+k_{i,exc_i}x_{exc,i}(t)
\end{equation}
where, $k_{p,exc}$ is the proportional gain and $k_{i,exc}$ is the integral gain of exciter. Similar to frequency measurements in governor, the voltage measurements are transmitted over the communication channel to the exciter to take the relevant control action. The adversary can therefore eavesdrop on these measurements to initiate the attack by manipulating the voltage data. The  derivation of the transient response of the rotor speed and the DC link voltage during these attacks is explained in the subsequent section.

\section{Transient Analysis During\\Data Integrity Attacks}
In order to derive the effect of data integrity attacks on the PGM governor and exciter, let us consider that there are $N$ synchronous generators, each integrated through 2L-VSC to the DC side in SPS. For the $i^{th}$ PGM, $i\in\{1,2,\hdots,N\}$, the parameters of the synchronous generator, converter, loads and network can be found in \cite{15}.

\subsection{Rotor Dynamics during Attack on the Governor}
The rotor angle dynamics of the synchronous generator can be represented as:
\begin{equation}
\label{eq4}
    \dot{\theta}_i(t)=\varphi \left ( \omega_i(t)-\omega_s \right )
\end{equation}
where, the time derivative of the rotor angle is represented by $\dot{\theta}$, the system frequency is denoted by $\varphi=2\pi f$ ($f$ = 50 or 60 Hz), the measured rotor speed (in rad/s) is represented by $\omega$, and the reference speed (i.e, synchronous speed) is represented by $\omega_s$.

Further, the rotor speed dynamics (in per unit -- p.u.) are represented as \cite{17}:
\begin{equation}
\label{eq5}
    \dot{\omega}_i(t)=\frac{\omega_s}{2H_i}\left ( P_{mi}(t)-P_{ei}(t)-D_i\left ( \omega_i(t)-\omega_s \right ) \right )
\end{equation}
where, $\dot{\omega}$ is the time derivative of the rotor speed, $H$ is the inertia constant, $P_m$ is the mechanical power, $P_e$ is the electrical power, and $D$ is the damping constant. The electrical power can also be represented in terms of generator voltage ($V$) and rotor angle ($\theta$) as:
\begin{equation}
\label{eq6}
    P_{ei}=\sum_{k=1}^{N}|V_i||V_k|(G_{ik}cos(\theta_i-\theta_k)+B_{ik}sin(\theta_i-\theta_k))
\end{equation}
where, $G_{ik}=G_{ki}$ and $B_{ik}=B_{ki}$ are the Kron-reduced equivalent conductance and susceptance between $i^{th}$ and $k^{th}$ synchronous generators, respectively. Eq. \eqref{eq5} represents the swing equation, which describes the electromechanical dynamics of the synchronous generator. This is useful to study the operation of the synchronous generators when subjected to a large disturbance (i.e., during transient periods). During normal operation, $P_m(t)=P_e(t)$. On the contrary, during a large disturbance, this equality cannot be satisfied, which may cause either increase or decrease in the rotor speed (depending on the inequality obtained in that duration). Further, a large deviation of rotor speed will impact the frequency of the system. If these fluctuations in frequency hit the threshold value of the protection devices, it may cause tripping of relays. This may both cause damage of the synchronous machine and disconnection of the machine affecting the reliability of electrical supply. 

In order to investigate the impact of large disturbance on the rotor speed, let us assume that the deviation in the rotor speed is represented as $\Delta \omega_i(t)=\omega_i(t)-\omega_s$.  Substituting this in Eq. \eqref{eq4} and swing equation Eq. \eqref{eq5}, we obtain:
\begin{equation}
\label{eq7}
\left\{ \begin{aligned}
  & \dot{ \theta}_i(t)=\varphi \Delta \omega_i(t) \\ 
 & \Delta \dot{\omega}_i(t)=\frac{\omega_s}{2H_i}\left ( P_{mi}(t)-P_{ei}(t)-D_i\Delta \omega_i(t) \right ) \\ 
\end{aligned} \right.
\end{equation}
In steady state, 
\begin{equation}
\label{eq8}
\left\{ \begin{aligned}
  & {\bar\theta}_i=\lim_{t\rightarrow \infty }\theta_i(t) \\ 
 & {\bar\omega}_i=\lim_{t\rightarrow \infty }\omega_i(t) \\ 
\end{aligned} \right.
\end{equation}
where, $\bar\theta$ and $\bar \omega$ are the steady state values of rotor angle and rotor speed, respectively. 

As mentioned, the SPS is transformed in a cyber-physical system, comprising of the communication links to exchange the measurements from the sensors periodically to the control units for the corresponding control action. In this case, the measurements of the rotor angle and rotor speed are communicated. Depending on the received measurements, the governor takes the control action to maintain the rotor speed to the desired value. To initiate the data integrity attack, the \emph{threat model} considers the following assumptions for the adversaries (as in \cite{23}): \textit{(i)} attackers can eavesdrop the sensor measurements, \textit{(ii)} they can also inject the manipulated measurements. The threat model can be justified due to the existing vulnerabilities of various communication protocols as mentioned earlier. Let $\hat{\omega}$ and $\hat{P_e}$ be the measured values of rotor speed and electrical power. Substituting these measured values, which are under attack (the $P_{e}$ is indirectly impacted due to the attack on rotor angle as shown in Eq. \eqref{eq6}) for $\Delta \dot{\omega}$ in Eq. \eqref{eq7}, we obtain:
\begin{equation}
\label{eq9}
    \Delta \dot{\omega}_i(t)=\frac{\omega_s}{2H_i}\left ( P_{mi}(t)-\hat{P}_{ei}(t)-D_i\Delta \hat{\omega}_i(t) \right )
\end{equation}

A generalized expression of the FDIAs on the different measurements can be represented as:
\begin{equation}
\label{eq10}
\left\{ \begin{aligned}
  & \Delta \hat{\omega}_i(t)=\Delta \omega_i(t)+\underbrace{\alpha_{i1}\Delta \omega_i(t)+\beta_{i1}(t)+\gamma_{i1}}_{\chi_{\omega_i}(t)} \\ 
 & \hat{P}_{ei}(t)=P_{ei}(t)+\underbrace{\alpha_{i2}P_{ei}(t)+\beta_{i2}(t)+\gamma_{i2}}_{\chi_{P_i}(t)} \\ 
\end{aligned} \right.
\end{equation}
where, $\chi_{\omega}$ is the fabricated measurement in the rotor speed, and $\chi_{P}$ is the fabricated measurement in the electrical power. The amplification term is represented by $\alpha$, the time-varying term is denoted by $\beta$, and the constant term is shown by $\gamma$. The subscripts 1, 2 represent the attack on $\Delta \omega$ and $P_e$, respectively. 

Substituting the fabricated measurements in Eq. \eqref{eq9}, we get:
\begin{equation}
\label{eq11}
\begin{aligned}
    \Delta \dot{\omega}_i(t)= & \frac{1}{2H_i/\omega_s}\{ -D_i\Delta \omega_i(t)-D_i\alpha_{i1}\Delta \omega_i(t)-\\ &\left ( D_i\gamma_{i1}+\gamma_{i2} \right )-( D_i \beta_{i1}(t)+\beta_{i2}(t)+\\ & \alpha_{i2}P_{ei}(t) ) \}
\end{aligned}
\end{equation}
Simplifying the above expression, we obtain:
\begin{equation}
\label{eq12}
    \Delta \dot{\omega}_i(t)=\underbrace{\lambda_{i1} \Delta \omega_i(t)}_{\text{no attack term}}+\underbrace{\lambda_{i2} \Delta \omega_i(t)+\lambda_{i3}+\lambda_{i4}(t)}_{\text{attack term}}
\end{equation}
where,
\begin{equation}
\label{eq13}
\left\{ \begin{aligned}
  & \lambda_{i1}=-\frac{D_i}{2H_i/\omega_s} \\ 
 & \lambda_{i2}=-\frac{D_i\alpha_{i1}}{2H_i/\omega_s} \\ 
 & \lambda_{i3}=-\frac{D_i\gamma_{i1}+\gamma_{i2}}{2H_i/\omega_s} \\
 & \lambda_{i4}(t)=-\frac{D_i\beta_{i1}(t)+\beta_{i2}(t)+\alpha_{i2}P_{ei}(t)}{2H_i/\omega_s} \\
\end{aligned} \right.
\end{equation}
Further, Eq. \eqref{eq12} is separated into a `no attack term' and an `attack term' to represent the impact of attack separately, i.e.: 
\begin{equation}
\label{eq14}
    \Delta \dot{\omega}_i(t)=\left ( \lambda_{i1}+\lambda_{i2} \right )\Delta \omega_i(t)+\lambda_{i3}+\lambda_{i4}(t)
\end{equation}
Solving this differential equation, we obtain:
\begin{equation}
\label{eq15}
\begin{aligned}
    \Delta \omega_i(t)=&\Delta \omega_i(t_o)e^{\left (\lambda_{i1}+\lambda_{i2}  \right )t}+\int_{t_o}^{t}\lambda_{i4}(\tau)e^{\left (\lambda_{i1}+\lambda_{i2}  \right )\left (t-\tau  \right )}d\tau\\ & +\frac{\lambda_{i3}}{\lambda_{i1}+\lambda_{i2}}\left ( e^{\left (\lambda_{i1}+\lambda_{i2}  \right )t}-1 \right )
\end{aligned}
\end{equation}
where, the term $t_o$ is the initial time, and $\Delta \omega_i(t_o)$ is the relative rotor speed at $t_o$. By relation of $\dot{\theta}$ and $\Delta \omega$ as expressed in Eq. \eqref{eq7}, we obtain the rotor angle response as:
\begin{equation}
\label{eq16}
    \begin{aligned}
        \theta_i(t)=&\theta_i(t_o)-\frac{\Delta \omega_i(t_o)(\lambda_{i1}+\lambda_{i2})+\lambda_{i3}}{(\lambda_{i1}+\lambda_{i2})^2}\varphi -\frac{\lambda_{i3}\varphi t}{\lambda_{i1}+\lambda_{i2}}+\\& \frac{\Delta \omega_i(t_o)(\lambda_{i1}+\lambda_{i2})+\lambda_{i3}}{(\lambda_{i1}+\lambda_{i2})^2}\varphi  e^{(\lambda_{i1}+\lambda_{i2})t} +\\ & \frac{\varphi}{\lambda_{i1}+\lambda_{i2}}\int_{t_o}^{t}\lambda_{i4}(\tau)(e^{(\lambda_{i1}+\lambda_{i2})(t-\tau)}-1)d\tau
    \end{aligned}
\end{equation}

The impact on rotor angle and rotor speed for various cases are discussed below.

\subsubsection{Case study 1 -- Normal operation without FDIAs}  
During normal operation when $\chi_{\omega_i}=0$ and $\chi_{P_i}=0$, substituting these values in Eq. \eqref{eq15}, we obtain the rotor speed:
    \begin{equation}
    \label{eq17}
    \Delta \omega_i(t)=\Delta \omega_i(t_o)e^{\lambda_{i1}t}
\end{equation}
Further, by Eq. \eqref{eq16}, the rotor angle obtained during normal operation is:
\begin{equation}
\label{eq18}
\theta_i(t)=\theta_i(t_o)-\frac{\varphi \Delta \omega_i(t_o)}{\lambda_{i1}}(1-e^{\lambda_{i1}t})
\end{equation}
It is to be noted that the term $\lambda_{i1}<0$, hence, the value of rotor speed and angle during steady state can be expressed as:
\begin{equation}
\label{eq19}
\left\{ \begin{aligned}
  & \Delta \bar{\omega}_i=0 \\ 
 & \bar{\theta}_i=\theta_i(t_o)-\frac{\varphi \Delta \omega_i(t_o)}{\lambda_{i1}} \\
\end{aligned} \right.
\end{equation}
This implies that the speed of the rotor will converge to the synchronous speed as deviation in rotor speed is zero. It means that the synchronous generator is operating in stable condition.

\subsubsection{Case study 2 -- Under constant bias FDIAs}
Consider an attack only on the rotor speed measurement ($\Delta \hat{\omega_i}$), i.e., $\chi_{\omega_i} \neq 0$ and $\chi_{P_i}=0$. We also consider a constant bias attack (i.e., $\gamma_{i1} \neq 0$ and all other terms are zero). During such scenario, the expressions for the rotor speed and angle can be obtained from Eqs. \eqref{eq15} -- \eqref{eq16} as:
\begin{equation}
\label{eq20}
    \Delta \omega_i(t)=\Delta \omega_i(t_o)e^{\lambda_{i1}t}+\frac{\gamma_{i1}}{D_i}(e^{\lambda_{i1}t}-1)
\end{equation}
\begin{equation}
\label{eq21}
\begin{aligned}
    \theta_i(t)=& \theta_i(t_o)+\frac{\varphi \Delta \omega_i(t_o)}{\lambda_{i1}}(1-e^{\lambda_{i1}t})- \\ & \frac{\gamma_{i1}\varphi }{D_i}(\lambda_{i1}+t)(1-e^{\lambda_{i1}t})
\end{aligned}
\end{equation}
In steady state, 
\begin{equation}
\label{eq22}
\left\{ \begin{aligned}
  & \Delta \bar{\omega}_i=-\frac{\gamma_{i1}}{D_i} \\ 
 & \bar{\theta}_i(t)=\theta_i(t_o)+\frac{\varphi \Delta \omega_i(t_o)}{\lambda_{i1}}-\frac{\gamma_{i1}\varphi }{D_i}(\lambda_{i1}+t) \\
\end{aligned} \right.
\end{equation}
It can be observed that the magnitude of the angle of rotor increases with time. 

\subsubsection{Case study 3 -- Under amplification FDIAs}
Again, we ssume the attack only on the rotor speed measurement ($\Delta \hat{\omega_i}$), i.e., $\chi_{\omega_i} \neq 0$ and $\chi_{P_i}=0$. Now, however, we consider an amplification attack (i.e., $\alpha_{i1} \neq 0$ and all other terms are zero). During such scenario, the expressions for the rotor speed and angle can obtained from Eqs. \eqref{eq15} -- \eqref{eq16} as:
\begin{equation}
\label{eq23}
    \Delta \omega_i(t)=\Delta \omega_i(t_o)e^{(\lambda_{i1}+\lambda_{i2})t}
\end{equation}
\begin{equation}
\label{eq24}
    \theta_i(t)=\theta_i(t_o)-\frac{\varphi \Delta \omega_i(t_o)}{\lambda_{i1}+\lambda_{i2}}(1+e^{(\lambda_{i1}+\lambda_{i2})t})
\end{equation}
For ($\lambda_{i1}+\lambda_{i2})<0$, the rotor speed and angle in steady state can be expressed as:
\begin{equation}
\label{eq25}
\left\{ \begin{aligned}
  & \Delta \bar{\omega}_i=0 \\ 
 & \bar{\theta}_i(t)=\theta_i(t_o)-\frac{\varphi \Delta \omega_i(t_o)}{\lambda_{i1}+\lambda_{i2}} \\
\end{aligned} \right.
\end{equation}
Hence, in this condition, the magnitude of rotor angle will increase. This will cause transient instability in the system.

\begin{figure*}[t]
    \centering
    \includegraphics[width=0.9\textwidth]{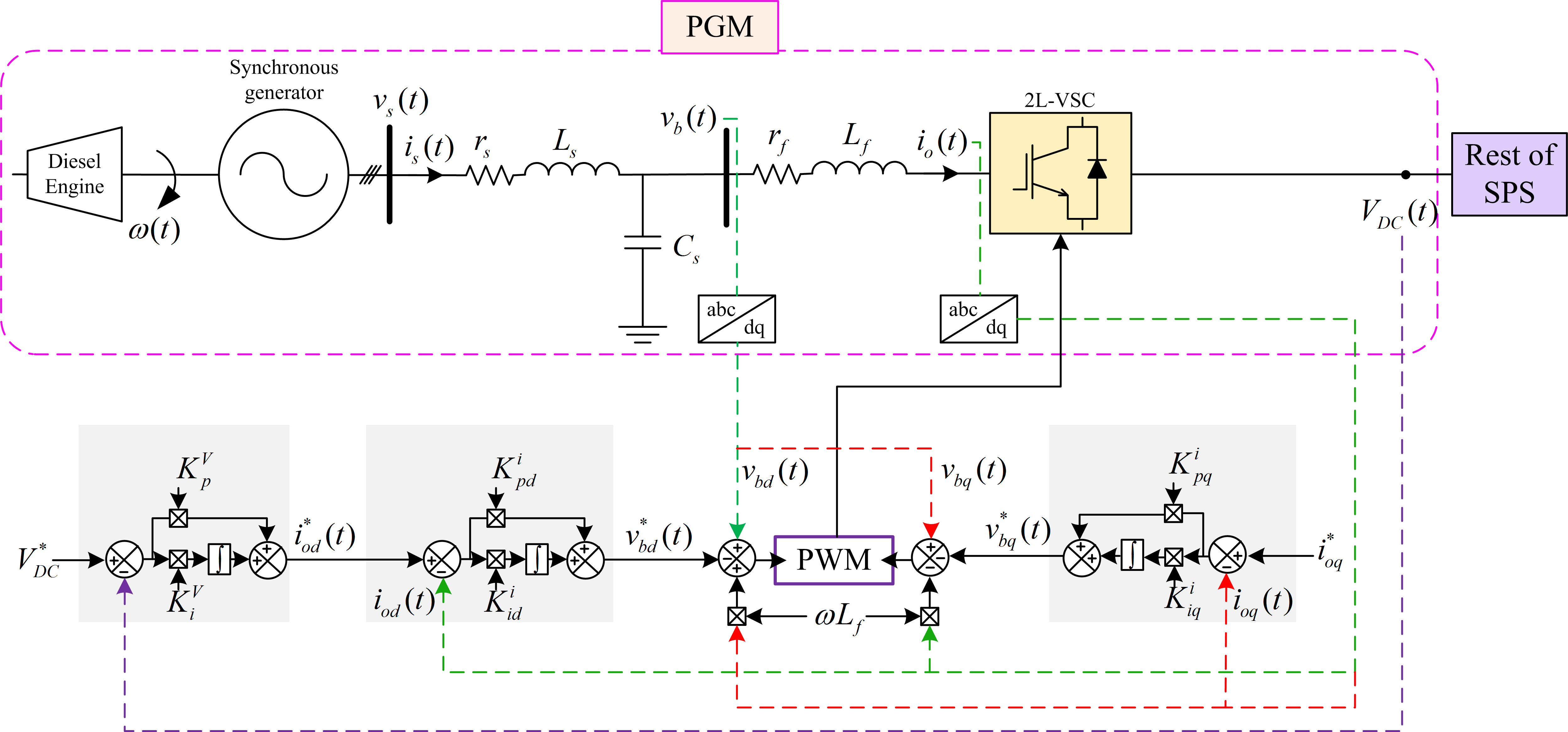}
    \caption{Voltage-source converter control architecture.}
    \label{fig:Rect}
\end{figure*}

\subsection{DC Link Voltage Dynamics during Attack on the Exciter}
The control architecture (in $dq0$ frame) of the converter for power conversion is represented in Fig. \ref{fig:Rect}. The dynamics of the source current ($i_s$) are represented as \cite{24}:
\begin{equation}
\label{eq26}
\left\{ \begin{aligned}
  & \dot{i}_{sd}(t)=\frac{1}{L_s}\left [-r_si_{sd}(t)+\omega L_si_{sq}(t)-v_{bd}(t)+v_{sd}(t)  \right ] \\ 
 & \dot{i}_{sq}(t)=\frac{1}{L_s}\left [ -\omega L_si_{sd}(t)-r_si_{sq}(t)-v_{bq}(t)+v_{sq}(t) \right ] \\
\end{aligned} \right.
\end{equation}
where, $r_s$ and $L_s$ represent the source resistance and inductance, respectively. Similarly, the bus voltage ($v_b$) dynamics are expressed as:
\begin{equation}
\label{eq27}
\left\{ \begin{aligned}
  & \dot{v}_{bd}(t)=\frac{1}{C_s}\left [ i_{sd}(t)+\omega C_sv_{bq}(t)-i_{od}(t) \right ] \\ 
 & \dot{v}_{bq}(t)=\frac{1}{C_s}\left [ i_{sq}(t)-\omega C_sv_{bd}(t)-i_{oq}(t) \right ] \\
\end{aligned} \right.
\end{equation}
where, $C_s$ denotes the source capacitance and $i_o$ represents the output current. Similarly, the dynamics of output current ($i_o$) are expressed as:
\begin{equation}
\label{eq28}
\left\{ \begin{aligned}
  & \dot{i}_{od}(t)=&\frac{1}{L_f}\left[-r_fi_{od}(t)+\underbrace{K_{pd}^i(i_{od}^*-i_{od}(t))+K_{id}^i\phi_d^i(t)}_\text{$v_{bd}^*$(t)}\right] \\ 
 & \dot{i}_{oq}(t)=&\frac{1}{L_f}\left[-r_fi_{oq}(t)+\underbrace{K_{pq}^i(i_{oq}^*-i_{oq}(t))+K_{iq}^i\phi_q^i(t)}_\text{$v_{bq}^*$(t)}\right] \\
\end{aligned} \right.
\end{equation}
where, $\dot{\phi}_d^i(t)= i_{od}^*(t)-i_{od}(t)$ and $\dot{\phi}_q^i(t)= i_{oq}^*-i_{oq}(t)$. Here, $r_f$ and $L_f$ represent the filter resistance and filter inductance, respectively. The superscript * represents the reference quantities. The proportional and integral gains of the current controller for $d-$axis are denoted by $K_{pd}^i$ and $K_{id}^i$, respectively. Similar terms can be obtained for $q-$axis as well. Further, the reference $d-$axis output current is expressed as:
\begin{equation}
\label{eq29}
    i_{od}^*(t)=K_{p}^V(V_{DC}^*-V_{DC}(t))+K_{i}^V \phi^V(t)
\end{equation}
where $\dot{\phi}^V(t)= V_{DC}^*-V_{DC}(t)$. The DC link voltage is represented as $V_{DC}$. The proportional and integral gains of the DC voltage regulator are denoted by $K_{p}^V$ and $K_{i}^V$, respectively. 

The DC link voltage dynamics can be expressed as:
\begin{equation}
\label{eq30}
    \dot{V}_{DC}(t)=\frac{1}{C_{DC}V_{DC}(t)}(\sum_{i=1}^{N}P_c(t)-P_l(t))
\end{equation}
where $C_{DC}$ is the capacitance on the DC network side. Further, $P_l(t)=V_{DC}(t)i_l(t)$ is the load power on the DC side, and $P_{c,i}(t)=\frac{3}{2}( i_{od,i}(t)(v_{bd,i}(t)-v_{bd,i}^*(t))+i_{oq,i}(t)(v_{bq,i}(t)-v_{bq,i}^*(t)) )$ is the power of $i^{th}$ converter. Substituting these expressions in Eq. \eqref{eq30}, we obtain: 
\begin{equation}
\label{eq31}
    \begin{aligned}
        \dot{V}_{DC}(t)=&\frac{3}{2C_{DC}V_{DC}(t)}(\sum_{i=1}^{N}(v_{bd,i}(t)i_{od,i}(t)+v_{bq,i}(t)i_{oq,i}(t) \\ & -K_{id}^ii_{od,i}(t)\phi_d^i(t)-K_{pd}^iK_i^Vi_{od,i}(t)\phi^V(t)+ \\ & K_{pd}^ii_{od,i}^2(t)-K_{pd}^iK_p^Vi_{od,i}(t)V_{DC}^*+K_{pq}^ii_{oq,i}^2(t)- \\ & K_{pq}^ii_{oq,i}(t)i_{oq,i}^*(t)-K_{iq}^ii_{oq,i}(t)\phi_q^i(t)+ \\ & K_{pd}^iK_p^Vi_{od,i}(t)V_{DC}(t))
-\frac{2}{3}i_lV_{DC}(t))
    \end{aligned}
\end{equation}
Let the measured bus voltage is represented as $\hat{v}_b$, then Eq. \eqref{eq31} can be rewritten as:
\begin{equation}
\label{eq32}
    \begin{aligned}
        \dot{V}_{DC}(t)=&\frac{3}{2C_{DC}V_{DC}(t)}(\sum_{i=1}^{N}(\hat{v}_{bd,i}(t)i_{od,i}(t)+\hat{v}_{bq,i}(t)i_{oq,i}(t)\\ & -K_{id}^ii_{od,i}(t)\phi_d^i(t)- K_{pd}^iK_i^Vi_{od,i}(T)\phi^V(t)+ \\ & K_{pd}^ii_{od,i}^2(t)-K_{pd}^iK_p^Vi_{od,i}(t)V_{DC}^*+K_{pq}^ii_{oq,i}^2(t)- \\ & K_{pq}^ii_{oq,i}(t)i_{oq,i}^*-K_{iq}^ii_{oq,i}(t)\phi_q^i(t)+\\ & K_{pd}^iK_p^Vi_{od,i}(t)V_{DC}(t))
-\frac{2}{3}i_l(t)V_{DC}(t))
    \end{aligned}
\end{equation}

Let the bus voltage measurements required by the exciter are under FDIAs as:
\begin{equation}
\label{eq33}
\left\{ \begin{aligned}
  & \hat{v}_{bd,i}(t)=v_{bd,i}(t)+\alpha_{i3}v_{bd,i}(t)+\beta_{i3}(t)+\gamma_{i3} \\ 
 & \hat{v}_{bq,i}(t)=v_{bq,i}(t)+\alpha_{i4}v_{bq,i}(t)+\beta_{i4}(t)+\gamma_{i4} \\
\end{aligned} \right.
\end{equation}
where, all terms have the meanings as defined before. The subscript 3, 4 represent attacks on $d-$axis and $q-$axis quantities of bus voltage, respectively. Including the measured bus voltage parameters in Eq. \eqref{eq32}, we get:
\begin{equation}
\label{eq34}
    \begin{aligned}
        \dot{V}_{DC}^A(t)=& \dot{V}_{DC}(t)+\frac{3}{2C_{DC}V_{DC}(t)}\sum_{i=1}^{N}((\alpha_{i3}v_{bd,i}(t)+ \\& \beta_{i3}(t)+ \gamma_{i3})i_{od,i}(t)+(\alpha_{i4}v_{bq,i}(t)+\beta_{i4}(t)+ \\& \gamma_{i4})i_{oq,i}(t))
    \end{aligned}
\end{equation}
where, the superscript A denotes the DC link voltage term under FDIA. Assuming the attack coefficients be positive, we obtain the deviation of DC link voltage will also be positive and vice-versa. Such a situation can cause the DC link voltage to trigger the threshold limits causing disconnection of the generator/loads, hence confirming the success of attack.

\subsection{Relation between Rotor Speed and DC Link Voltage}
We have investigated the impact on rotor speed and DC link voltage individually due to the data integrity on the governor and exciter respectively. Here, we derive the relationship between the deviation in rotor speed and the deviation in DC link voltage. On the AC side of the $i^{th}$ converter, the swing equation can be expressed as:
\begin{equation}
\label{eq35}
    \frac{2H_i}{\omega_s}\dot{\omega_i}(t)=\Delta P_{ACi}(t)
\end{equation}
where, $\Delta P_{ACi}(t)=P_{mi}(t)-P_{ei}(t)$. On the DC side of the $i^{th}$ converter, the power equation can be expressed as:
\begin{equation}
\label{eq36}
    \frac{C_{DC}V_{DC}(t)}{S_c}\dot{V}_{DC}(t)=\Delta P_{DC}(t)
\end{equation}
where, $S_c$ is the converter power rating, and $\Delta P_{DC}$ is the deviation in power in the DC side. Further, for power balance within the SPS, equating Eq. \eqref{eq35} and Eq. \eqref{eq36}, we obtain:
\begin{equation}
\label{eq37}
    \frac{4H_iS_c}{\omega_sC_{DC}}(\omega_i(t)-\omega_s)=V_{DC}^2(t)-V_{DC}^{*2}
\end{equation}
Let $\Delta V_{DC}(t)=V_{DC}(t)-V_{DC}^*$, we obtain:
\begin{equation}
\label{eq38}
    \Delta \omega_i(t)=\frac{C_{DC}\omega_s}{4H_iS_c}\left \{ \left (\Delta V_{DC}(t) + V_{DC}^* \right )^2 -V_{DC}^* \right \}
\end{equation}
Hence, the deviation in rotor speed and DC link voltage holds a nonlinear relationship. The positive change in DC link voltage causes a positive change in rotor speed and vice-versa, which coincides with the physical characteristics of the system.

\begin{figure*}[t]
\centering
	\begin{subfigure}[h!]{0.32\textwidth}
	\centering
	\includegraphics[width=1\linewidth]{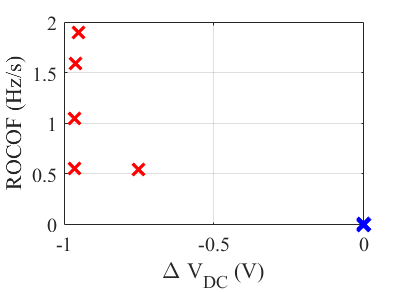}
		\caption{}
		\label{fig:F}
		\vspace{0.2cm}
	\end{subfigure}
	\hfill
	\begin{subfigure}[h!]{0.32\textwidth}
	\centering
	\includegraphics[width=1\linewidth]{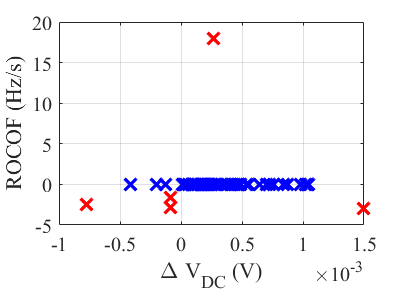}
	\caption{}
	\label{fig:A1}
	\vspace{0.2cm}
	\end{subfigure}
        \hfill
	\begin{subfigure}[h!]{0.32\textwidth}
	\centering
	\includegraphics[width=1\linewidth]{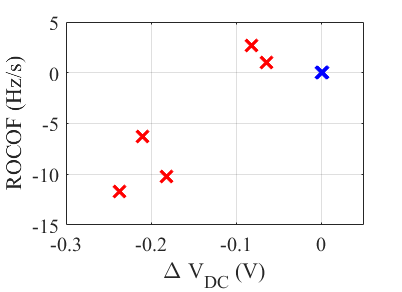}
	\caption{}
	\label{fig:A2}
	\vspace{0.2cm}
	\end{subfigure}
\vspace{-2mm}	
 \caption{Phase portrait between ROCOF and deviation in DC link voltage for ({a}) fault; ({b}) data integrity attack on governor and~({c}) data integrity attack on exciter of PGM 1.}
	\label{fig:C}
\end{figure*}

\begin{figure*}[t]
\centering
	\begin{subfigure}[h!]{0.32\textwidth}
	\centering
	\includegraphics[width=1\linewidth]{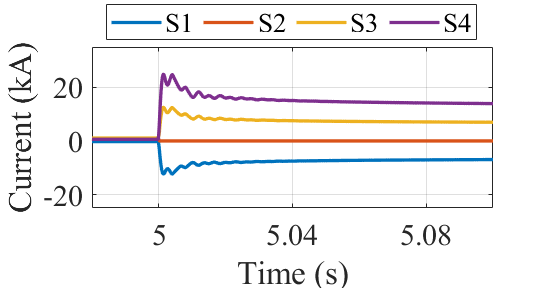}
		\caption{}
		\label{fig:f1}
		\vspace{0.2cm}
	\end{subfigure}
	\hfill
	\begin{subfigure}[h!]{0.32\textwidth}
	\centering
	\includegraphics[width=1\linewidth]{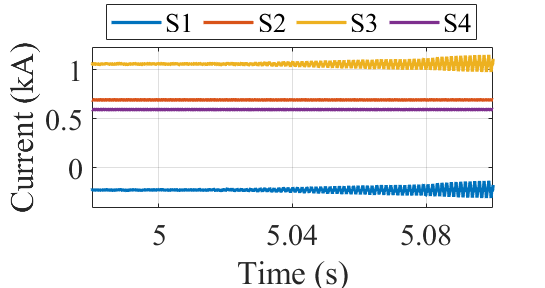}
	\caption{}
	\label{fig:a1}
	\vspace{0.2cm}
	\end{subfigure}
        \hfill
	\begin{subfigure}[h!]{0.32\textwidth}
	\centering
	\includegraphics[width=1\linewidth]{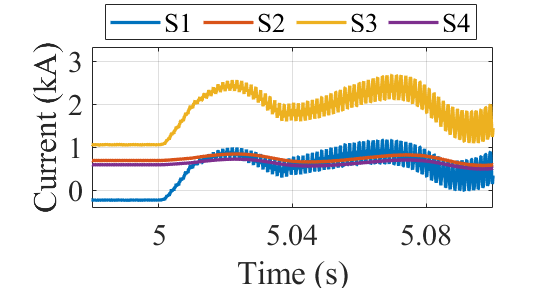}
	\caption{}
	\label{fig:a2}
	\vspace{0.2cm}
	\end{subfigure}
\vspace{-2mm}	
 \caption{Current in circuit breakers A, B, C, D for ({a}) fault; ({b}) data integrity attack of low value on governor and~({c}) data integrity attack of low value on exciter of PGM 1.}
	\label{fig:D}
\end{figure*}

\begin{figure*}[t]
\centering
	\begin{subfigure}[h!]{0.32\textwidth}
	\centering
	\includegraphics[width=1\linewidth]{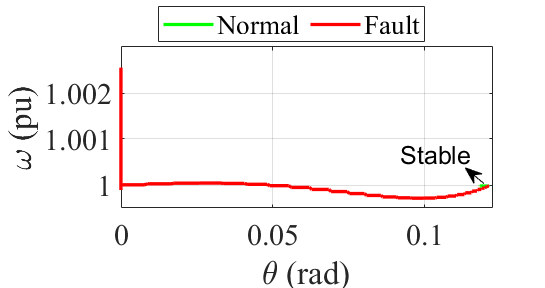}
		\caption{}
		\label{fig:f1A}
		\vspace{0.2cm}
	\end{subfigure}
	\hfill
	\begin{subfigure}[h!]{0.32\textwidth}
	\centering
	\includegraphics[width=1\linewidth]{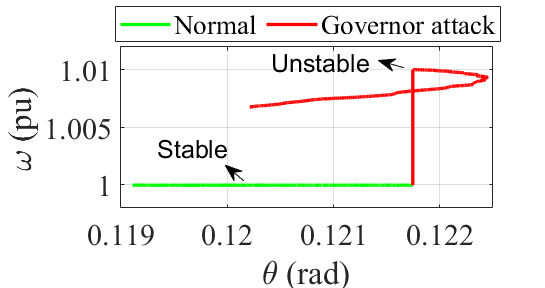}
	\caption{}
	\label{fig:a1A}
	\vspace{0.2cm}
	\end{subfigure}
        \hfill
	\begin{subfigure}[h!]{0.32\textwidth}
	\centering
	\includegraphics[width=1\linewidth]{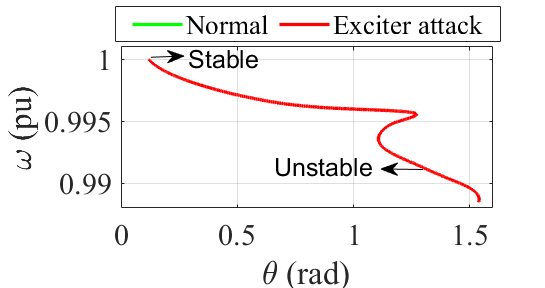}
	\caption{}
	\label{fig:a2A}
	\vspace{0.2cm}
	\end{subfigure}
\vspace{-2mm}	
 \caption{Phase-portrait of PGM 1 in $\omega-\theta$ domain during ({a}) fault; ({b}) data integrity attack of low value on governor and~({c}) data integrity attack of low value on exciter of PGM 1.}
	\label{fig:Angle}
\end{figure*}

\section{Result and Discussion}
The impact of FDIAs on the considered 12 kV MVDC SPS is discussed in this section. These attacks can have similar consequences (on the protection devices) as that of a large disturbance (i.e., fault) able to cause tripping of the protection devices resulting in disconnection of source/loads.

\subsection{Fault Event}
During large disturbances, such as faults, the installed protection devices may operate to isolate the faulty section and protect the healthy part of the system. Therefore, for the protection of power systems against transient frequency response, several relays, such as under/over frequency relays and rate of change of frequency (ROCOF) relays, are installed. The allowable deviation in frequency is $\pm$5 \% from the nominal value. Moreover, the acceptable value for ROCOF relay is $\pm$0.02 Hz/s \cite{25}. During fast changes in frequency, the ROCOF relay will trip if it crosses the threshold value (even though the frequency is within the allowable range). Moreover, for the protection of power systems against transient voltage response, the over/under voltage relays are deployed. According to IEC 60092-101 guidelines, the continuous voltage deviations for DC distribution systems should be within $\pm$ 10\% \cite{26}. If the circuit breakers are not disconnected within the allowed critical time during faults, the system might become unstable. A phase portrait of the ROCOF and the deviation in DC link voltage for MVDC SPS during fault event is shown in Fig. \ref{fig:F}. Here, the blue cross represents the pre-disturbance samples and the red cross represent the samples during the disturbance (i.e., fault). It can be observed that during fault, the samples lie outside the allowed threshold limits. Hence, this may cause disconnection of the generator/loads if the fault is not isolated with the stipulated time. The currents in the switches (S1, S2, S3 and S4) during fault are shown in Fig. \ref{fig:f1}. Moreover, the phase-portrait of PGM 1 in $\omega-\theta$ domain during fault is shown in Fig. \ref{fig:Angle}(a). 

\subsection{Data Integrity Attacks}
Although the attack can result in sub-optimal operating conditions, in this work, the attack is considered successful if it results in tripping relays and causing network reduction (due to disconnection of sources/loads). Let us consider the attack on the governor and exciter of PGM 1. A phase portrait of the ROCOF and the deviation in DC link voltage for MVDC SPS during data integrity attack on governor is shown in Fig. \ref{fig:A1}. Similar to the data integrity attack on the governor, a phase portrait of the ROCOF and the deviation in DC link voltage for MVDC SPS during attack on exciter is shown in Fig. \ref{fig:A2}. It can be observed that during both attacks, the samples lie outside the allowed threshold limits, hence confirming the successful attack. Thus, the protection devices will trip and govern the circuit breakers to cause local/full shutdown. This would affect the reliability and continuity of electrical supply. The currents in the switches (S1, S2, S3 and S4) are shown in Figs. \ref{fig:a1} and \ref{fig:a2}. It can be observed that for data integrity attack of low value on governor, the currents show oscillatory response, which may cause the system instability. Similarly for data integrity attack of low value on exciter, the system tends to become unstable as well. Moreover, the phase-portrait of PGM 1 in $\omega-\theta$ domain during data integrity attack on governor and exciter are shown in Figs. \ref{fig:Angle}(b) and \ref{fig:Angle}(c). With the acceptable variations in frequency being $\pm 0.5\%$ of the nominal value, it can be seen that with the injection of data integrity attack, the system tends towards instability. 

\section{Conclusions and Future Work}
The work in this paper discusses the impact of falsified attack signals affecting the data integrity through FDIAs on the PGMs on the SPS. An analysis of such attacks has been carried out on the governor and exciter control of the synchronous generator (part of PGM). The system under the study is a 12 kV MVDC nominal two-zone SPS simulated in MATLAB/Simulink. The transient performance of the SPS due to such attacks has been investigated on the rotor speed and DC link voltage parameters. The relation between the deviation in rotor speed and the deviation in DC link voltage has been also derived. This work provides an impact assessment outcome for data integrity attacks on SPS. It can further be extended to analyze the impact of cyber-attacks on SPS due to compromise of different cyber elements in the network (e.g., switches, radio frequency communication channels, etc.).

{\bibliographystyle{IEEEtran}

@ARTICLE{1,
  author={Jin, Zheming and Sulligoi, Giorgio and Cuzner, Rob and Meng, Lexuan and Vasquez, Juan C. and Guerrero, Josep M.},
  journal={IEEE Electrific. Mag.}, 
  title={Next-Generation Shipboard {DC} Power System: Introduction Smart Grid and {DC} Microgrid Technologies into Maritime Electrical Netowrks}, 
  year={2016},
  volume={4},
  number={2},
  pages={45-57},
  doi={10.1109/MELE.2016.2544203}}

  @ARTICLE{2,
  author={Sulligoi, Giorgio and Bosich, Daniele and Vicenzutti, Andrea and Khersonsky, Yuri},
  journal={IEEE Trans. Ind. Appl}, 
  title={Design of Zonal Electrical Distribution Systems for Ships and Oil Platforms: Control Systems and Protections}, 
  year={2020},
  volume={56},
  number={5},
  pages={5656-5669},
  doi={10.1109/TIA.2020.2999035}}

@ARTICLE{3,
  author={},
  journal={IEEE Std. 1826-2012}, 
  title={{IEEE Power Electronics Open System Interfaces in Zonal Electrical Distribution Systems Rated Above 100 kW}}, 
  year={2012},
  volume={},
  number={},
  pages={},
  doi={}}

@ARTICLE{4,
  author={},
  journal={IEEE Std. 1676-2010}, 
  title={{IEEE Guide for Control Architecture for High Power Electronics (1 MW and Greater) Used in Electric Power Transmission and Distribution Systems}}, 
  year={2011},
  volume={},
  number={},
  pages={},
  doi={}}

@misc{5,
author = {C. Cimpanu},
title = {All four of the world's largest shipping companies have now been hit by cyber-attacks},
year = {2020 (accessed February 11, 2023)}, 
howpublished = "\url{https://www.zdnet.com/article/all-four-of-the-worlds-largest-shipping-companies-have-now-been-hit-by-cyber-attacks/}"}

@ARTICLE{6,
  author={F. Roberts and D. Egan and C. Nelson and R. Whytlaw},
  journal={NMIOTC Marit. Interdiction Oper. J.}, 
  title={Combined  cyber and physical attacks on the maritime transportation  system}, 
  year={2019},
  volume={18},
  number={22},
  pages={},
  doi={}}

  @misc{7,
author = {{The Maritime Executive}},
title = {Cyberattack Hits Multiple Greek Shipping Firms},
year = {2021 (accessed February 11, 2023)}, 
howpublished = "\url{https://www.maritime-executive.com/article/cyberattack-hits-multiple-greek-shipping-firms}"}

@misc{8,
author = {H. Cooper},
title = {Chinese Hackers Steal Unclassified Data from Navy Contractor},
year = {2018 (accessed February 11, 2023)}, 
howpublished = "\url{https://www.nytimes.com/2018/06/08/us/politics/china-hack-navy-contractor-.html}"}

@ARTICLE{9,
  author={Sahoo, Subham and Dragičević, Tomislav and Blaabjerg, Frede},
  journal={IEEE J. Emerg. Sel. Topics Power Electron.}, 
  title={Cyber Security in Control of Grid-Tied Power Electronic Converters—Challenges and Vulnerabilities}, 
  year={2021},
  volume={9},
  number={5},
  pages={5326-5340},
  doi={10.1109/JESTPE.2019.2953480}}

@ARTICLE{27,
  author={Anubi, Olugbenga Moses and Konstantinou, Charalambos},
  journal={IEEE Trans. Ind. Informat.}, 
  title={Enhanced Resilient State Estimation Using Data-Driven Auxiliary Models}, 
  year={2020},
  volume={16},
  number={1},
  pages={639-647},
  doi={10.1109/TII.2019.2924246}}

@ARTICLE{10,
  author={Xiao, Zhao-Xia and Li, Huai-Min and Fang, Hong-Wei and Guan, Yu-Zhe and Liu, Tao and Hou, Lucas and Guerrero, Josep M.},
  journal={IEEE Trans. Transp. Electrific.}, 
  title={Operation Control for Improving Energy Efficiency of Shipboard Microgrid Including Bow Thrusters and Hybrid Energy Storages}, 
  year={2020},
  volume={6},
  number={2},
  pages={856-868},
  doi={10.1109/TTE.2020.2992735}}

@ARTICLE{11,
  author={Park, Daeseong and Zadeh, Mehdi},
  journal={IEEE Trans. Transp. Electrific}, 
  title={Modeling and Predictive Control of Shipboard Hybrid DC Power Systems}, 
  year={2021},
  volume={7},
  number={2},
  pages={892-904},
  doi={10.1109/TTE.2020.3027184}}

@ARTICLE{12,
  author={Wang, Yu and Mondal, Suman and Satpathi, Kuntal and Xu, Yan and Dasgupta, Souvik and Gupta, Amit Kumar},
  journal={IEEE Trans. Transp. Electrific.}, 
  title={Multiagent Distributed Power Management of {DC} Shipboard Power Systems for Optimal Fuel Efficiency}, 
  year={2021},
  volume={7},
  number={4},
  pages={3050-3061},
  doi={10.1109/TTE.2021.3086303}}

@ARTICLE{13,
  author={M. A. Ben Farah and E. Ukwandu and H. Hindy and D. Brosset and M. Bures and
I. Andonovic and X. Bellekens},
  journal={Information}, 
  title={Cyber Security in the Maritime Industry: A Systematic Survey of Recent Advances and Future Trends}, 
  year={2022},
  volume={13},
  number={1},
  pages={},
  doi={}}

@ARTICLE{14,
  author={K. Gupta and S. Sahoo and B. K. Panigrahi and F. Blaabjerg and P. Popovski},
  journal={Energies}, 
  title={On the Assessment of Cyber Risks and Attack Surfaces in a Real- Time Co-Simulation Cybersecurity Testbed for Inverter-Based Microgrids}, 
  year={2021},
  volume={14},
  number={16},
  pages={4941},
  doi={https://www.mdpi.com/1996-1073/14/16/4941}}

  @ARTICLE{15,
  author={N. Doerry},
  journal={Corbin A. McNeill Symposium, United States Naval Academy, Annapolis}, 
  title={Next Generation Integrated Power Systems for the Future Fleet}, 
  year={2009},
  volume={},
  number={},
  pages={},
  doi={}}

  @misc{16,
author = {{U.S. Department of Homeland Security}},
title = {Northern California Area Maritime Security Committee Cyber Security News Letter},
year = {2022 (accessed February 11, 2023)}, 
howpublished = "\url{https://www.sfmx.org/wp-content/uploads/2022-07_Cyber-Newsletter.pdf}"}

@book{17,
  title     = "Power System Dynamics: Stability and Control",
  author    = "J. Machowski and Z. Lubosny and J. W. Bialek and J. R. Bumby",
  year      = 2020,
  publisher = "Hoboken, NJ, USA: Wiley",
  url       = ""
}

@ARTICLE{18,
  author={Almas, Muhammad Shoaib and Vanfretti, Luigi},
  journal={IEEE Access}, 
  title={A Hybrid Synchrophasor and {GOOSE}-Based Power System Synchronization Scheme}, 
  year={2016},
  volume={4},
  number={},
  pages={4659-4668},
  doi={10.1109/ACCESS.2016.2601445}}

@ARTICLE{19,
  author={Best, Robert J. and Morrow, D. John and McGowan, David J. and Crossley, Peter A.},
  journal={IEEE Trans.Power Systems}, 
  title={Synchronous Islanded Operation of a Diesel Generator}, 
  year={2007},
  volume={22},
  number={4},
  pages={2170-2176},
  doi={10.1109/TPWRS.2007.907449}}

  @ARTICLE{20,
  author={Wang, Yu and Mondal, Suman and Satpathi, Kuntal and Xu, Yan and Dasgupta, Souvik and Gupta, Amit Kumar},
  journal={IEEE Trans. Transp. Electrific.}, 
  title={Multiagent Distributed Power Management of DC Shipboard Power Systems for Optimal Fuel Efficiency}, 
  year={2021},
  volume={7},
  number={4},
  pages={3050-3061},
  doi={10.1109/TTE.2021.3086303}}

@ARTICLE{21,
  author={Roomi, Muhammad M. and Hussain, S. M. Suhail and Mashima, Daisuke and Chang, Ee-Chien and Ustun, Taha Selim},
  journal={IEEE Sys. J.}, 
  title={Analysis of False Data Injection Attacks Against Automated Control for Parallel Generators in {IEC 61850}-Based Smart Grid Systems}, 
  year={2023},
  volume={},
  number={},
  pages={1-12},
  doi={10.1109/JSYST.2023.3236951}}

@ARTICLE{22,
  author={Jo, Hyo Jin and Choi, Wonsuk},
  journal={IEEE Trans. Intell. Transp. Sys.}, 
  title={A Survey of Attacks on Controller Area Networks and Corresponding Countermeasures}, 
  year={2022},
  volume={23},
  number={7},
  pages={6123-6141},
  doi={10.1109/TITS.2021.3078740}}


@inproceedings{23,
  title={CHIMERA: A Hybrid Estimation Approach to Limit the Effects of False Data Injection Attacks},
  author={Liu, Xiaorui and Hu, Yaodan and Konstantinou, Charalambos and Jin, Yier},
  booktitle={2021 IEEE International Conference on Communications, Control, and Computing Technologies for Smart Grids (SmartGridComm)},
  pages={95--101},
  year={2021},
  organization={IEEE}
}

  @ARTICLE{24,
  author={Pogaku, Nagaraju and Prodanovic, Milan and Green, Timothy C.},
  journal={IEEE Trans. Power Electron.}, 
  title={Modeling, Analysis and Testing of Autonomous Operation of an Inverter-Based Microgrid}, 
  year={2007},
  volume={22},
  number={2},
  pages={613-625},
  doi={10.1109/TPEL.2006.890003}}

  @ARTICLE{25,
  author={T. Kerdphol and Y. Matsukawa and M. Watanabe and Y. Mitani},
  journal={Electr. Power Syst. Res.}, 
  title={Application of {PMUs} to monitor large-scale {PV} penetration infeed on frequency of 60 {Hz} {Japan} power system: {A} case study from {Kyushu Island}}, 
  year={2020},
  volume={185},
  number={1},
  pages={Art. no. 106393},
  doi={}}

@ARTICLE{26,
  author={Park, Daeseong and Zadeh, Mehdi},
  journal={IEEE Trans. Transp. Electrific.}, 
  title={Modeling and Predictive Control of Shipboard Hybrid DC Power Systems}, 
  year={2021},
  volume={7},
  number={2},
  pages={892-904},
  doi={10.1109/TTE.2020.3027184}}

@inproceedings{ogilvie2020modeling,
  title={Modeling communication networks in a real-time simulation environment for evaluating controls of shipboard power systems},
  author={Ogilvie, Colin and Ospina, Juan and Konstantinou, Charalambos and Vu, Tuyen and Stanovich, Mark and Schoder, Karl and Steurer, Mischa},
  booktitle={IEEE CyberPELS},
  pages={1--7},
  year={2020},
  organization={IEEE}
}

@INPROCEEDINGS{9512343,
  author={Ospina, Juan and Konstantinou, Charalambos and Stanovich, Mark and Steurer, Mischa},
  booktitle={2021 IEEE Electric Ship Technologies Symposium (ESTS)}, 
  title={Evaluation of Communication Network Models for Shipboard Power Systems}, 
  year={2021},
  volume={},
  number={},
  pages={1-9},
  doi={10.1109/ESTS49166.2021.9512343}}

@INPROCEEDINGS{9512317,
  author={Nguyen, Bang L. H. and Vu, Tuyen and Ogilvie, Colin and Ravindra, Harsha and Stanovich, Mark and Schoder, Karl and Steurer, Michael and Konstantinou, Charalambos and Ginn, Herbert and Schegan, Christian},
  booktitle={2021 IEEE Electric Ship Technologies Symposium (ESTS)}, 
  title={Advanced Load Shedding for Integrated Power and Energy Systems}, 
  year={2021},
  volume={},
  number={},
  pages={1-6},
  doi={10.1109/ESTS49166.2021.9512317}}


\begin{thebibliography}{10}
\providecommand{\url}[1]{#1}
\csname url@samestyle\endcsname
\providecommand{\newblock}{\relax}
\providecommand{\bibinfo}[2]{#2}
\providecommand{\BIBentrySTDinterwordspacing}{\spaceskip=0pt\relax}
\providecommand{\BIBentryALTinterwordstretchfactor}{4}
\providecommand{\BIBentryALTinterwordspacing}{\spaceskip=\fontdimen2\font plus
\BIBentryALTinterwordstretchfactor\fontdimen3\font minus
  \fontdimen4\font\relax}
\providecommand{\BIBforeignlanguage}[2]{{%
\expandafter\ifx\csname l@#1\endcsname\relax
\typeout{** WARNING: IEEEtran.bst: No hyphenation pattern has been}%
\typeout{** loaded for the language `#1'. Using the pattern for}%
\typeout{** the default language instead.}%
\else
\language=\csname l@#1\endcsname
\fi
#2}}
\providecommand{\BIBdecl}{\relax}
\BIBdecl

\bibitem{1}
Z.~Jin, G.~Sulligoi, R.~Cuzner, L.~Meng, J.~C. Vasquez, and J.~M. Guerrero,
  ``Next-generation shipboard {DC} power system: Introduction smart grid and
  {DC} microgrid technologies into maritime electrical netowrks,'' \emph{IEEE
  Electrific. Mag.}, vol.~4, no.~2, pp. 45--57, 2016.

\bibitem{2}
G.~Sulligoi, D.~Bosich, A.~Vicenzutti, and Y.~Khersonsky, ``Design of zonal
  electrical distribution systems for ships and oil platforms: Control systems
  and protections,'' \emph{IEEE Trans. Ind. Appl}, vol.~56, no.~5, pp.
  5656--5669, 2020.

\bibitem{3}
``{IEEE Power Electronics Open System Interfaces in Zonal Electrical
  Distribution Systems Rated Above 100 kW},'' \emph{IEEE Std. 1826-2012}, 2012.

\bibitem{4}
``{IEEE Guide for Control Architecture for High Power Electronics (1 MW and
  Greater) Used in Electric Power Transmission and Distribution Systems},''
  \emph{IEEE Std. 1676-2010}, 2011.

\bibitem{5}
C.~Cimpanu, ``All four of the world's largest shipping companies have now been
  hit by cyber-attacks,''
  \url{https://www.zdnet.com/article/all-four-of-the-worlds-largest-shipping-companies-have-now-been-hit-by-cyber-attacks/},
  2020 (accessed February 11, 2023).

\bibitem{6}
F.~Roberts, D.~Egan, C.~Nelson, and R.~Whytlaw, ``Combined cyber and physical
  attacks on the maritime transportation system,'' \emph{NMIOTC Marit.
  Interdiction Oper. J.}, vol.~18, no.~22, 2019.

\bibitem{7}
{The Maritime Executive}, ``Cyberattack hits multiple greek shipping firms,''
  \url{https://www.maritime-executive.com/article/cyberattack-hits-multiple-greek-shipping-firms},
  2021 (accessed February 11, 2023).

\bibitem{8}
H.~Cooper, ``Chinese hackers steal unclassified data from navy contractor,''
  \url{https://www.nytimes.com/2018/06/08/us/politics/china-hack-navy-contractor-.html},
  2018 (accessed February 11, 2023).

\bibitem{9}
S.~Sahoo, T.~Dragičević, and F.~Blaabjerg, ``Cyber security in control of
  grid-tied power electronic converters—challenges and vulnerabilities,''
  \emph{IEEE J. Emerg. Sel. Topics Power Electron.}, vol.~9, no.~5, pp.
  5326--5340, 2021.

\bibitem{27}
O.~M. Anubi and C.~Konstantinou, ``Enhanced resilient state estimation using
  data-driven auxiliary models,'' \emph{IEEE Trans. Ind. Informat.}, vol.~16,
  no.~1, pp. 639--647, 2020.

\bibitem{10}
Z.-X. Xiao, H.-M. Li, H.-W. Fang, Y.-Z. Guan, T.~Liu, L.~Hou, and J.~M.
  Guerrero, ``Operation control for improving energy efficiency of shipboard
  microgrid including bow thrusters and hybrid energy storages,'' \emph{IEEE
  Trans. Transp. Electrific.}, vol.~6, no.~2, pp. 856--868, 2020.

\bibitem{11}
D.~Park and M.~Zadeh, ``Modeling and predictive control of shipboard hybrid dc
  power systems,'' \emph{IEEE Trans. Transp. Electrific}, vol.~7, no.~2, pp.
  892--904, 2021.

\bibitem{12}
Y.~Wang, S.~Mondal, K.~Satpathi, Y.~Xu, S.~Dasgupta, and A.~K. Gupta,
  ``Multiagent distributed power management of {DC} shipboard power systems for
  optimal fuel efficiency,'' \emph{IEEE Trans. Transp. Electrific.}, vol.~7,
  no.~4, pp. 3050--3061, 2021.

\bibitem{ogilvie2020modeling}
C.~Ogilvie, J.~Ospina, C.~Konstantinou, T.~Vu, M.~Stanovich, K.~Schoder, and
  M.~Steurer, ``Modeling communication networks in a real-time simulation
  environment for evaluating controls of shipboard power systems,'' in
  \emph{IEEE CyberPELS}.\hskip 1em plus 0.5em minus 0.4em\relax IEEE, 2020, pp.
  1--7.

\bibitem{9512343}
J.~Ospina, C.~Konstantinou, M.~Stanovich, and M.~Steurer, ``Evaluation of
  communication network models for shipboard power systems,'' in \emph{2021
  IEEE Electric Ship Technologies Symposium (ESTS)}, 2021, pp. 1--9.

\bibitem{13}
M.~A.~B. Farah, E.~Ukwandu, H.~Hindy, D.~Brosset, M.~Bures, I.~Andonovic, and
  X.~Bellekens, ``Cyber security in the maritime industry: A systematic survey
  of recent advances and future trends,'' \emph{Information}, vol.~13, no.~1,
  2022.

\bibitem{14}
K.~Gupta, S.~Sahoo, B.~K. Panigrahi, F.~Blaabjerg, and P.~Popovski, ``On the
  assessment of cyber risks and attack surfaces in a real- time co-simulation
  cybersecurity testbed for inverter-based microgrids,'' \emph{Energies},
  vol.~14, no.~16, p. 4941, 2021.

\bibitem{9512317}
B.~L.~H. Nguyen, T.~Vu, C.~Ogilvie, H.~Ravindra, M.~Stanovich, K.~Schoder,
  M.~Steurer, C.~Konstantinou, H.~Ginn, and C.~Schegan, ``Advanced load
  shedding for integrated power and energy systems,'' in \emph{2021 IEEE
  Electric Ship Technologies Symposium (ESTS)}, 2021, pp. 1--6.

\bibitem{15}
N.~Doerry, ``Next generation integrated power systems for the future fleet,''
  \emph{Corbin A. McNeill Symposium, United States Naval Academy, Annapolis},
  2009.

\bibitem{16}
{U.S. Department of Homeland Security}, ``Northern california area maritime
  security committee cyber security news letter,''
  \url{https://www.sfmx.org/wp-content/uploads/2022-07_Cyber-Newsletter.pdf},
  2022 (accessed February 11, 2023).

\bibitem{17}
J.~Machowski, Z.~Lubosny, J.~W. Bialek, and J.~R. Bumby, \emph{Power System
  Dynamics: Stability and Control}.\hskip 1em plus 0.5em minus 0.4em\relax
  Hoboken, NJ, USA: Wiley, 2020.

\bibitem{18}
M.~S. Almas and L.~Vanfretti, ``A hybrid synchrophasor and {GOOSE}-based power
  system synchronization scheme,'' \emph{IEEE Access}, vol.~4, pp. 4659--4668,
  2016.

\bibitem{19}
R.~J. Best, D.~J. Morrow, D.~J. McGowan, and P.~A. Crossley, ``Synchronous
  islanded operation of a diesel generator,'' \emph{IEEE Trans.Power Systems},
  vol.~22, no.~4, pp. 2170--2176, 2007.

\bibitem{20}
Y.~Wang, S.~Mondal, K.~Satpathi, Y.~Xu, S.~Dasgupta, and A.~K. Gupta,
  ``Multiagent distributed power management of dc shipboard power systems for
  optimal fuel efficiency,'' \emph{IEEE Trans. Transp. Electrific.}, vol.~7,
  no.~4, pp. 3050--3061, 2021.

\bibitem{21}
M.~M. Roomi, S.~M.~S. Hussain, D.~Mashima, E.-C. Chang, and T.~S. Ustun,
  ``Analysis of false data injection attacks against automated control for
  parallel generators in {IEC 61850}-based smart grid systems,'' \emph{IEEE
  Sys. J.}, pp. 1--12, 2023.

\bibitem{22}
H.~J. Jo and W.~Choi, ``A survey of attacks on controller area networks and
  corresponding countermeasures,'' \emph{IEEE Trans. Intell. Transp. Sys.},
  vol.~23, no.~7, pp. 6123--6141, 2022.

\bibitem{23}
X.~Liu, Y.~Hu, C.~Konstantinou, and Y.~Jin, ``Chimera: A hybrid estimation
  approach to limit the effects of false data injection attacks,'' in
  \emph{2021 IEEE International Conference on Communications, Control, and
  Computing Technologies for Smart Grids (SmartGridComm)}.\hskip 1em plus 0.5em
  minus 0.4em\relax IEEE, 2021, pp. 95--101.

\bibitem{24}
N.~Pogaku, M.~Prodanovic, and T.~C. Green, ``Modeling, analysis and testing of
  autonomous operation of an inverter-based microgrid,'' \emph{IEEE Trans.
  Power Electron.}, vol.~22, no.~2, pp. 613--625, 2007.

\bibitem{25}
T.~Kerdphol, Y.~Matsukawa, M.~Watanabe, and Y.~Mitani, ``Application of {PMUs}
  to monitor large-scale {PV} penetration infeed on frequency of 60 {Hz}
  {Japan} power system: {A} case study from {Kyushu Island},'' \emph{Electr.
  Power Syst. Res.}, vol. 185, no.~1, p. Art. no. 106393, 2020.

\bibitem{26}
D.~Park and M.~Zadeh, ``Modeling and predictive control of shipboard hybrid dc
  power systems,'' \emph{IEEE Trans. Transp. Electrific.}, vol.~7, no.~2, pp.
  892--904, 2021.

\end{thebibliography}

}

\end{document}